\documentclass[sn-mathphys,numbered]{sn-jnl}

\usepackage{graphicx}%
\usepackage{multirow}%
\usepackage{amsmath,amssymb,amsfonts}%
\usepackage{amsthm}%
\usepackage{mathrsfs}%
\usepackage[title]{appendix}%
\usepackage{xcolor}%
\usepackage{textcomp}%
\usepackage{manyfoot}%
\usepackage{booktabs}%
\usepackage{algorithm}%
\usepackage{algorithmicx}%
\usepackage{algpseudocode}%
\usepackage{listings}%

\usepackage{subcaption}

\theoremstyle{thmstyleone}%
\theoremstyle{thmstyletwo}%

\theoremstyle{thmstylethree}%

\newcommand{\B}       [1] {{\bf {#1}}}

\newcommand{\derp}[2]{\partial #1/ \partial #2}

\newcommand{\Derp}[2]{\frac{ \partial #1 }{ \partial #2 }}

\newcommand {\mb}[1]{\mathbf{#1}}
 
\newcommand{\Res}{\mb{R}} 
\newcommand{\Jac}{\mb{A}}

\newcommand{\Mf}{\mb{M}_\mathrm{F}}
\newcommand{\Mr}{\mb{M}_\mathrm{R}}

\newcommand{\xr}{x_\mathrm{r}}
\newcommand{\rr}{r_\mathrm{r}}
\newcommand{\xf}{x_\mathrm{f}}
\newcommand{\rf}{r_\mathrm{f}}

\newcommand{\vbold}{\mb{v}}
\newcommand{\ubold}{\mb{u}}
\newcommand{\qbold}{\mb{q}}
\newcommand{\qbar}{\mb{\bar{q}}}
\newcommand{\qprime}{\mb{q'}}
\newcommand{\fbold}{\mb{f}}

\newcommand{\qhat}{\mb{\hat{q}}}
\newcommand{\fhat}{\mb{\hat{f}}}

\newcommand{\MatV}{\mb{V}}
\newcommand{\MatU}{\mb{U}}

\newcommand{\zbold}{\mb{z}}

\newcommand{\MatFlim}{\mb{B}}
\newcommand{\Resmod}{\mb{H}}

\newcommand{\Pqq}{\mb{P}_{qq}} 
\newcommand{\Pff}{\mb{P}_{ff}} 
\newcommand{\Pzz}{\mb{P}_{zz}}

\newcommand{\Nrmodes}{n_{\mathrm{res}}} 

\newcommand{\Echu}{E_{\mathrm{Chu}}}

\newcommand{\cphi}{c_\varphi}

\usepackage{xcolor}

\definecolor{darkGreen}{RGB}{4, 87, 1}
\definecolor{persoRed}{RGB}{232, 9, 9}

\begin{document}

\title[Acoustic resolvent analysis of turbulent jets]{Acoustic resolvent analysis of turbulent jets}










\author*[1,2]{\fnm{Benjamin} \sur{Bugeat}}\email{bb283@le.ac.uk}

\author[3]{\fnm{Ugur} \sur{Karban}}

\author[2]{\fnm{Anurag} \sur{Agarwal}}

\author[4]{\fnm{Lutz} \sur{Lesshafft}}

\author[5]{\fnm{Peter} \sur{Jordan}}

\affil*[1]{\orgdiv{School of Engineering}, \orgname{University of Leicester},  \city{Leicester}, \country{UK}}

\affil[2]{\orgdiv{Department of Engineering}, \orgname{University of Cambridge},  \city{Cambridge}, \country{UK}}

\affil[3]{\orgdiv{Department of Aerospace Engineering}, \orgname{Middle East Technical University},  \city{Ankara}, \country{Turkey}}

\affil[4]{\orgdiv{Laboratoire d'Hydrodynamique}, \orgname{CNRS,  \'{E}cole polytechnique, Institut Polytechnique de Paris},  \city{Palaiseau}, \country{France}}

\affil[5]{\orgdiv{D\'{e}partement Fluides, Thermique, Combustion}, \orgname{Institut Pprime, CNRS-University of Poitiers-ENSMA},  \city{Poitiers}, \country{France}}




\abstract{
We perform a resolvent analysis of a compressible turbulent jet, where the optimisation domain of the response modes is located in the acoustic field, excluding the hydrodynamic region, in order to promote acoustically efficient modes.
We examine the properties of the acoustic resolvent and assess its potential for jet-noise modelling, focusing on the subsonic regime.
Resolvent forcing modes, consistent with previous studies, are found to contain supersonic waves associated with Mach wave radiation in the response modes.
This differs from the standard resolvent in which hydrodynamic instabilities dominate.
We compare resolvent modes with SPOD modes educed from LES data.
Acoustic resolvent response modes generally have better alignment with acoustic SPOD modes than standard resolvent response modes.
For the optimal mode, the angle of the acoustic beam is close to that found in SPOD modes for moderate frequencies.
However, there is no significant separation between the singular values of the leading and sub-optimal modes.
Some suboptimal modes are furthermore shown to contain irrelevant structure for jet noise.
Thus, even though it contains essential acoustic features absent from the standard resolvent approach, the SVD of the acoustic resolvent alone is insufficient to educe a low-rank model for jet noise.
But because it identifies the prevailing mechanisms of jet noise, it provides valuable guidelines in the search of a forcing model (Karban \textit{et al.} An empirical model of noise sources in subsonic jets. \textit{Journal of Fluid Mechanics} (2023): A18).
}




\maketitle

\section{Introduction}


Jet-noise modelling, despite several decades of study, is still an active field of research.
Recent progress in LES computations have made possible accurate comparisons with experiments \citep{bres2012towards,bres2018importance}.
But robust low-rank models, useful for extracting the essential physical mechanisms at play as well as being an efficient tool for industry, remain an important research objective.
The apparently disorganised turbulent jet flow contains coherent structures \citep{mollo1967jet,crow1971orderly}.
These structures, termed wavepackets as they experience growth and decay within the turbulent mixing layer, play a central role in jet noise \citep{jordan2013wave,cavalieri2019wave}.
Modelling efforts have therefore been directed to wavepackets, using different variations of local stability  analysis -- parallel \citep{michalke1984survey}, WKB \citep{crighton1976stability}, PSE \citep{gudmundsson2011instability,cavalieri2013wavepackets} or OWNS \citep{rigas2017one}. 
Such approaches are sufficient to recover a large part of the acoustic field in supersonic jets \citep{sinha2014wavepacket}.
This is because Mach wave radiation, in which a supersonic disturbance generates an acoustic radiation, is the dominant mechanism of jet noise in this regime \citep{tam1995supersonic}.
As stability analysis provides a good model of the single-point, second-order statistics of these supersonic coherent structures, it is sufficient to deduce the associated, coupled acoustic wave.
Subsonic jet noise, however, requires higher-order statistical information, and therefore further modelling.
In this regime, the phase speed of coherent structures is subsonic.
Only a small fraction of the energy of wavepackets -- which is acoustically matched \citep{crighton1975basic}, is responsible for acoustic radiation.
Moreover, this acoustic radiation is greatly affected by the space-time modulation of wavepackets, termed jitter \citep{cavalieri2011jittering,cavalieri2014coherence}, and linked to the stochastic nature of the turbulent flow.
By cancelling destructive interferences, jitter may enhance sound pressure levels up to 30 dB \citep{cavalieri2014coherence,jordan2014modeling}.
As a result, information of the two-point statistics of these structures is required, which, in the spectral domain, appears in the cross-spectral density (CSD) matrix.

Resolvent analysis has recently gained popularity in the turbulence modelling community.
In this approach, an input-output view of the flow dynamics is adopted \citep{mckeon2010critical,hwang2010amplification}.
A linearised Navier-Stokes operator, through its selective receptivity to forcing input, produces a flow response that is likely to be statistically prevalent in the turbulent dynamics.
This forcing contains the non-linear interactions in the flow.
In the presence of strong amplification mechanisms, its modelling may not be required to obtain coherent structures \citep{beneddine2016conditions}.
In this case, the singular value decomposition (SVD) of the resolvent operator exhibits a large separation between the first and second singular values, and the first left singular vector alone can provide a good model for coherent structures.

Resolvent analysis was used to extract hydrodynamic wavepackets in turbulent jets \citep{schmidt2018spectral,lesshafft2019resolvent,pickering2021optimal}.
This allowed different mechanisms to be identified in the frequency range associated with jet noise: the Orr mechanism was found to be dominant at low frequency while Kelvin-Helmholtz wavepackets prevail at larger frequencies.
\cite{semeraro2016stochastic} and \cite{towne2016toward} furthermore noticed that resolvent analysis could be extended to a stochastic framework, opening perspectives for the inclusion of jitter.
In the stochastic framework, a correspondence is identified between resolvent modes and spectral proper orthogonal decomposition (SPOD) modes \citep{towne2018spectral} in the case where the forcing field is spatially uncorrelated. 
While this ``white-noise forcing'' hypothesis is not \emph{a priori} valid in turbulent flow, it has been found to successfully model the coherent dynamics in a range of different configurations, including flow over a backward-facing step \citep{beneddine2016conditions}, certain flames \citep{kaiser2019prediction} and, importantly, jets (as cited above), at least over a significant range of frequencies. The cited examples share a strong dominance of the Kelvin--Helmholtz instability in free shear layers. Many counter-examples exist as well, where the nonlinear forcing of the linear resolvent operator strongly excites sub-optimal response patterns, and the white-noise hypothesis is therefore unsuccessful in reproducing the SPOD spectrum. This is notably often observed in wall-bounded shear flows \citep[e.g.][]{morra2019relevance,morra2021colour,symon2021energy}.

\cite{cavalieri2019wave} pointed out fundamental similarities between resolvent analysis and acoustic analogies \citep{lighthill1952sound,goldstein2003generalized}.
Indeed, the resolvent operator stems from a reorganisation of the Navier-Stokes equations, resulting in a linear relation between a forcing and a response; in aeroacoustics, hydrodynamic sources are similarly associated with an acoustic field, and linked via a linear operator.
Several authors \citep{garnaud2013global,jeun2016input,pickering2021resolvent} used a modified version of resolvent analysis that more closely resembles acoustic analogies. 
Energy norms associated with the left and right singular vectors (the response and forcing fields, respectively) can be defined in the SVD of the resolvent operator \citep{sipp2013characterization}.
By restricting the response norm to the acoustic field -- rather than the whole numerical domain -- the SVD is expected to yield resolvent modes that are acoustically efficient.
This framework will be termed acoustic resolvent in this paper, as opposed to standard resolvent which includes both the hydrodynamic and acoustic regions in the response norm.

The study of \cite{garnaud2013global}, via the acoustic resolvent, was carried out on a subsonic turbulent jet at Reynolds number $Re \simeq 10^{3}$. 
The numerical domain included a nozzle, where the forcing field was constrained to be localised.
Good agreements were found with direct numerical simulation (DNS).
The authors noted, however, that most of the acoustic radiation was produced at the nozzle exit, hindering interpretations regarding jet noise.
\cite{jeun2016input} more extensively analysed the acoustic resolvent framework in turbulent jets at $Re = 10^6$, ranging from subsonic to supersonic Mach numbers.
The SVD was presented, including tens of singular values, showing that a clear separation of more than one order of magnitude existed between the first and the second ones in supersonic jets.
In subsonic jets, however, singular values were all found to be of the same order of magnitude.
The authors concluded on the need to incorporate more than one mode in jet-noise modelling through acoustic resolvent analysis -- even in supersonic jets, where subsequent modes were reported to participate in side-line noise.
\cite{pickering2021resolvent} recently used the acoustic resolvent to propose a novel modelling approach of jet noise. 
In order to model the second-order statistics of the acoustic field, the authors proposed a low-rank model based on the SVD of the acoustic resolvent, enriched by a model describing the projection of the forcing field onto the right singular vectors involving two calibration constants.
Encouraging results were obtained for the jet-noise prediction at low angles compared to LES data at $Re \simeq 10^6$.

The aim of the present paper is to examine the properties of the acoustic resolvent and to assess its potential for jet-noise modelling, focusing on the subsonic regime.
In particular, we aim at identifying and providing a physical interpretation of the fundamental differences between the standard and acoustic resolvent approaches.
Emphasis will be put on the physical nature of the resolvent forcing modes in order to get new insights into the modelling of jet-noise sources.
We will discuss the structure and phase speed of theses modes, which have been overlooked in the subsonic regime so far.
Unphysical modes, not reported in the literature, will furthermore be shown to arise, and hinder the use of the acoustic resolvent.
Comparisons with SPOD modes extracted from LES calculations will provide valuable insights to this end.
The influence of the nozzle, which can greatly affect jet noise \citep{bres2018importance,kaplan2021nozzle}, will also be investigated; previous studies on acoustic resolvent \citep{garnaud2013global,jeun2016input,pickering2021resolvent} did not discuss it.
Overall, the strengths, but also the limitations of the acoustic resolvent approach, will be highlighted.
Mean flows obtained from high-fidelity LES calculations at $Re \simeq 10^6$ will be used to carry out the resolvent analysis.

The paper is organised as follows.
The theoretical and numerical framework are introduced in section \ref{sec.resolvent}.
Afterwards, a detailed comparison of the SVD of the standard and acoustic resolvents is presented \ref{sec.standard_acoustic}.
The effect of the nozzle is analysed in section \ref{sec.nozzle}.
The acoustic resolvent modes are then more specifically studied in section \ref{sec.acousticResolvent}.
We examine their physical features, their alignment with SPOD modes and the effect of the Mach number.
A discussion about the physical interpretation of the SVD of the acoustic resolvent and its potential for jet-noise modelling follows in section \ref{sec.discussion}, before concluding the paper in section \ref{sec.conclusion}.


\section{Resolvent analysis} \label{sec.resolvent}

\subsection{Mean flow} 
\begin{figure}
    \centering
    \includegraphics[angle=-0,trim=0 0 0 0, clip,width=0.85\textwidth]{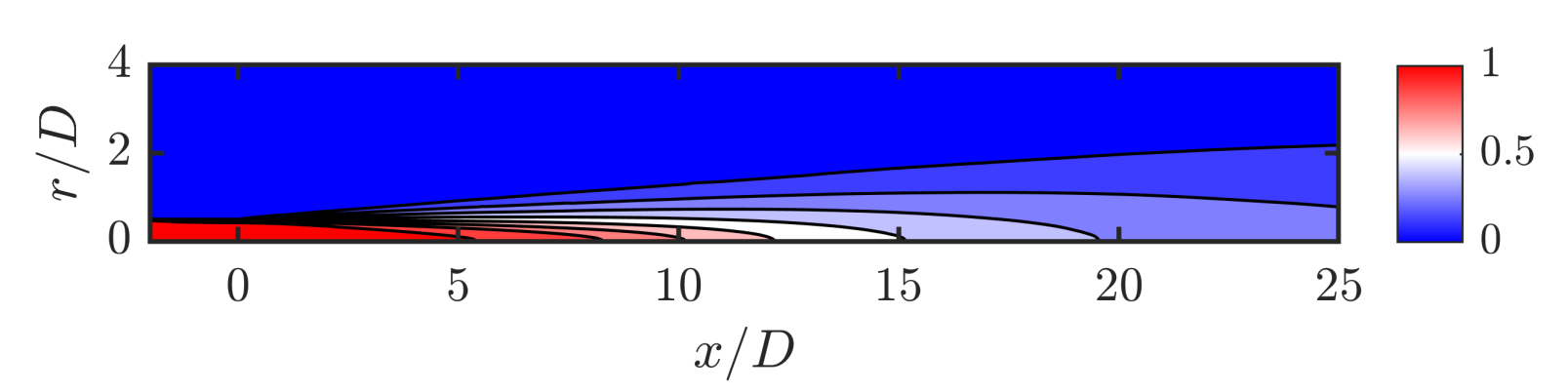}  
    \caption{Mean streamwise velocity at $M=0.9$ and $Re=1.01 \times 10^6$.}
     \label{fig.meanflow}
\end{figure}

Resolvent analysis involves the linearisation of the governing equations around a mean flow.
We use a database obtained from LES calculations using the code 'Charles'; details about the numerical procedure and the experimental validation can be found in \cite{bres2018importance}. 
In the present work, a statistically steady and axisymmetric turbulent jet flow is considered (figure \ref{fig.meanflow}).
The streamwise and radial directions are noted $x$ and $r$, respectively.
Note that a nozzle flow is included in the numerical domain.
The state vector $\qbold$ is expressed with the primitive variables $\qbold = ( \nu \> \> \mb{u} \> \> p)^t$. 
Here, $\nu$ is the specific volume, $\mb{u}$ the velocity vector in cylindrical coordinates and $p$ the pressure.  
All physical quantities are made dimensionless using the jet diameter $D$ and the density and velocity at the nozzle exit along the centre line $\rho_{\mathrm{j}}$ and $U_\mathrm{j}$. 
In the reference case, the Mach and Reynolds numbers are $M= U_\mathrm{j}/c_\mathrm{j}=0.9$ and $Re=\rho_\mathrm{j} U_\mathrm{j}D / \mu_\mathrm{j} = 1.01 \times 10^6$, respectively, where $c$ is the speed of sound and $\mu$ is the molecular viscosity.
The Strouhal number will be defined as $St = f D / U_\mathrm{j}$. 

\subsection{Resolvent matrix}

The discretised compressible non-linear Navier-Stokes equations are written as the following dynamical system:
\begin{align}\label{eq.NS}
\Derp{\mb{q}}{t} = \mb{N}(\mb{q}) .
\end{align}
Introducing the classical Reynolds decomposition $\mb{q}(\mb{x},t) = \qbar(\mb{x}) + \qprime(\mb{x},t)$, the governing equation of the perturbations reads
\begin{align}\label{eq.q_temp}
\Derp{\qprime}{t} = \Jac  \qprime + \mb{f} .
\end{align}

\noindent The right-hand term has been decomposed such that the Jacobian matrix calculated around the mean flow $\qbar$, $\Jac = \derp{\mathcal{N}}{\mb{q}}|_{\qbar}$, appears.
This constitutes the linear part of the equation while $\mb{f}$ contains the remaining non-linear terms. 
The analysis is now carried out in the frequency domain and restricted to \emph{axisymmetric} two-dimensional perturbations; the azimuthal wavenumber is always zero and does not enter the problem as a parameter. 
The governing equation becomes
\begin{align}\label{eq.qfreq}
i \omega \qhat = \Jac  \qhat + \fhat ,
\end{align}

\noindent where $\omega$ is the angular frequency .
Following the modelling approach of turbulent flows proposed by \cite{mckeon2010critical} and \cite{hwang2010amplification}, the term $\fhat$ represents the intrinsic turbulent background forcing originating from the coupling of different space and time scales.
Then, $\qhat$ can be seen as the linear response to this forcing.
Because $\qhat$ contains both energy amplification mechanisms in the hydrodynamic and acoustic regions, this approach can be used for jet-noise modelling. 
One can further note a strong resemblance with acoustic analogies \citep{lighthill1952sound,goldstein2003generalized}. 
The resolvent matrix can finally be introduced as the link, in the frequency domain, between the forcing and the response fields. 
Dropping the hat notations for the rest of the paper, the following relation holds:
\begin{align}\label{eq.res}
\qbold = \Res \fbold ,
\end{align}

\noindent where $\Res = \left( i \omega \mathbf{I} - \Jac \right)^{-1}$ and $\mathbf{I}$ is the identity matrix.

\subsection{Singular value decomposition}

The resolvent matrix contains all the information needed to compute the response to a given forcing in the frequency domain.
A singular value decomposition (SVD) can be performed in order to find a low-rank approximation of this matrix.
From a physical point of view, this amounts to the definition of a gain between the energy of the response and that of the forcing and an identification of the forcing fields that generate the largest values of this gain. 
The motivation behind the approach is to explore if, with only a few modes, a majority of the fluctuation of the system can be recovered; in other words, building a low-order model.
The approach detailed hereafter introduces a tailored resolvent framework to study jet noise \citep{garnaud2013global,jeun2016input,pickering2021resolvent}.
A measurement matrix $\MatFlim$ is introduced, such that  
\begin{align}
    \fbold &= \MatFlim \zbold \label{eq.reduced_variables2}.
\end{align}
This allows the forcing field to be restricted to a specified region and/or specific components of the forcing vector.
The modified resolvent matrix $\Resmod = \Res \MatFlim$ provides a link between the two new variables:
\begin{align}\label{eq.res_modified}
\qbold = \Resmod \zbold .
\end{align}
The gain between the energy of the response and the forcing, noted $\sigma^2$, is now defined as
\begin{align}\label{eq.gain}
    \sigma^2 = \frac{\langle \qbold,\qbold \rangle_{\mathrm{R}}}{\langle\zbold,\zbold\rangle_{\mathrm{F}}} .
\end{align}
Two scalar products have been introduced.
In a discrete framework, they are defined from weight matrices such that $\langle\qbold,\qbold\rangle_{\mathrm{R}} = \qbold^* \Mr \qbold$ and $\langle \zbold,\zbold \rangle_{\mathrm{F}} = \zbold^* \Mf \zbold$.
The choice of the norms is detailed in section \ref{sec.norms}.
Note that $\Mf$ has to be a positive-definite matrix.
This is not necessary for $\Mr$, which can be positive semi-definite \citep{garnaud2013global}.
From equations \eqref{eq.res_modified} and \eqref{eq.gain}, the gain can then be recast as a generalised Rayleigh product:
\begin{align}\label{eq.gainStep2}
    \sigma^2 =  \frac{\zbold^* \Resmod^* \Mr \Resmod \zbold}{\zbold^* \Mf  \zbold} .
\end{align}
\noindent We now aim at computing the forcing vectors that generate the largest (or optimal) gain.
Following \cite{sipp2013characterization}, this problem can be solved as a generalised SVD, which reads
\begin{align}\label{eq.gsvd2}
    \Resmod &= \MatU \mb{\Sigma} \MatV^* \Mf ,\\
    \MatU^* \Mr \MatU & = \MatV^* \Mf \MatV = \mathbf{I} .
\end{align}
The matrices $\MatV$ and $\MatU$ contain the right-singular and left-singular vectors $\vbold_i$ and $\ubold_i$, respectively.
The vectors $\vbold_i$ correspond to the optimal and sub-optimal forcing modes generating the optimal and sub-optimal gains $\sigma_1 > \sigma_2 > ...$ (equation \eqref{eq.gain}), which are contained in the diagonal matrix $\mb{\Sigma}$.
The optimal and sub-optimal response modes, associated with these forcing modes, are the vectors $\ubold_i$, which are readily computed using ${\MatU = \Resmod \MatV \mb{\Sigma}^{-1}}$.

\subsection{Low-rank approximation of the response}

Adopting a stochastic approach is relevant to deal with turbulent flows.
The second-order statistics of the system can be modelled within the resolvent framework by introducing the cross-spectral density (CSD) matrices of the response and the forcing $\Pqq$ and $\Pzz$, respectively.
\cite{semeraro2016stochastic} and \cite{towne2016toward} showed that
\begin{align}\label{eq.PqqPff}
    \Pqq = \Resmod \Pzz \Resmod^* .
\end{align}
Combined with the SVD of the resolvent in equation \eqref{eq.gsvd2}, $\Pqq$ can be expressed as
\begin{align}\label{eq.PqqLowrank}
    \Pqq = \mb{U} \mb{\Sigma} \mb{V}^* \Mf \Pzz \Mf^* \mb{V} \mb{\Sigma} \mb{U}^*.
\end{align}
An interesting simplification occurs under the assumption $(\Mf^{1/2})^* \Pff \Mf^{1/2} = \mathbb{I}$.
When $\Mf = \mathbb{I}$, this assumption means that the forcing field is being modelled as white noise in space \citep{semeraro2016stochastic}.
In this case, because of the orthonormality properties of the SVD, equation \eqref{eq.PqqLowrank} can be rewritten as
\begin{align}\label{eq.PqqWhiteNoise}
    \Pqq = \mb{U} \mb{\Sigma}^2 \mb{U}^* ,
\end{align}
which shows that resolvent modes are, in this case, identical to SPOD modes \citep{towne2018spectral}.
A low-rank approximation of the PSD of the state vector - which lies on the diagonal of $\Pqq$, can easily be computed as
\begin{align}\label{eq.PSDwhitenoise}
    PSD(\mb{q}) \simeq \sum_{k=1}^{\Nrmodes} \sigma_k^2 | \mb{u}_k |^2 .
\end{align}

\noindent  where $\Nrmodes$ the number of resolvent modes retained, i.e. the order of the truncation of the resolvent basis used to build the response.

\subsection{Energy measure} \label{sec.norms}

A measure of the energy of the forcing and response fields has to be chosen and implemented through matrices $\Mf$ and $\Mr$.
In compressible flows, it is common to use the energy measure introduced by \cite{chu1965energy}, which provides a measure of the energy of compressible disturbances that is free of non-physical growth, as shown by \cite{george2011chu}.
This energy can be derived for the primitive variables considered in our study by using the linear relationships between variables, which yields
\begin{equation} \label{eq.ChuDefinition}
    \Echu = \frac{1}{2} \int_{\Omega} \left( \frac{1}{\bar{\nu}} |\mb{u}|^2 + \frac{\gamma \bar{p}}{(\gamma-1)\bar{\nu}^2} \nu^2  +  \frac{2}{(\gamma-1)\bar{\nu}} \nu p + \frac{1}{(\gamma-1) \bar{p}} p^2 \right) r \mathrm{d}r \mathrm{d}x .
\end{equation}  
The integration domain $\Omega$, in the physical space $(x,r)$, can be chosen in order to compute the energy of the response in a specific portion of the numerical domain, thereby setting the definition of the gain in equation \eqref{eq.gain}.
As for the energy of the forcing, $\Omega$ is chosen such that it matches the size of vector $\zbold$ in order to ensure that the matrix $\Mf$ is positive-definite.

\subsection{Standard and acoustic resolvents: set-up} \label{sec.standard_acoustic}

\begin{figure}
    \centering    
        \includegraphics[angle=-0,trim=0 0 0 0, clip,width=0.85\textwidth]{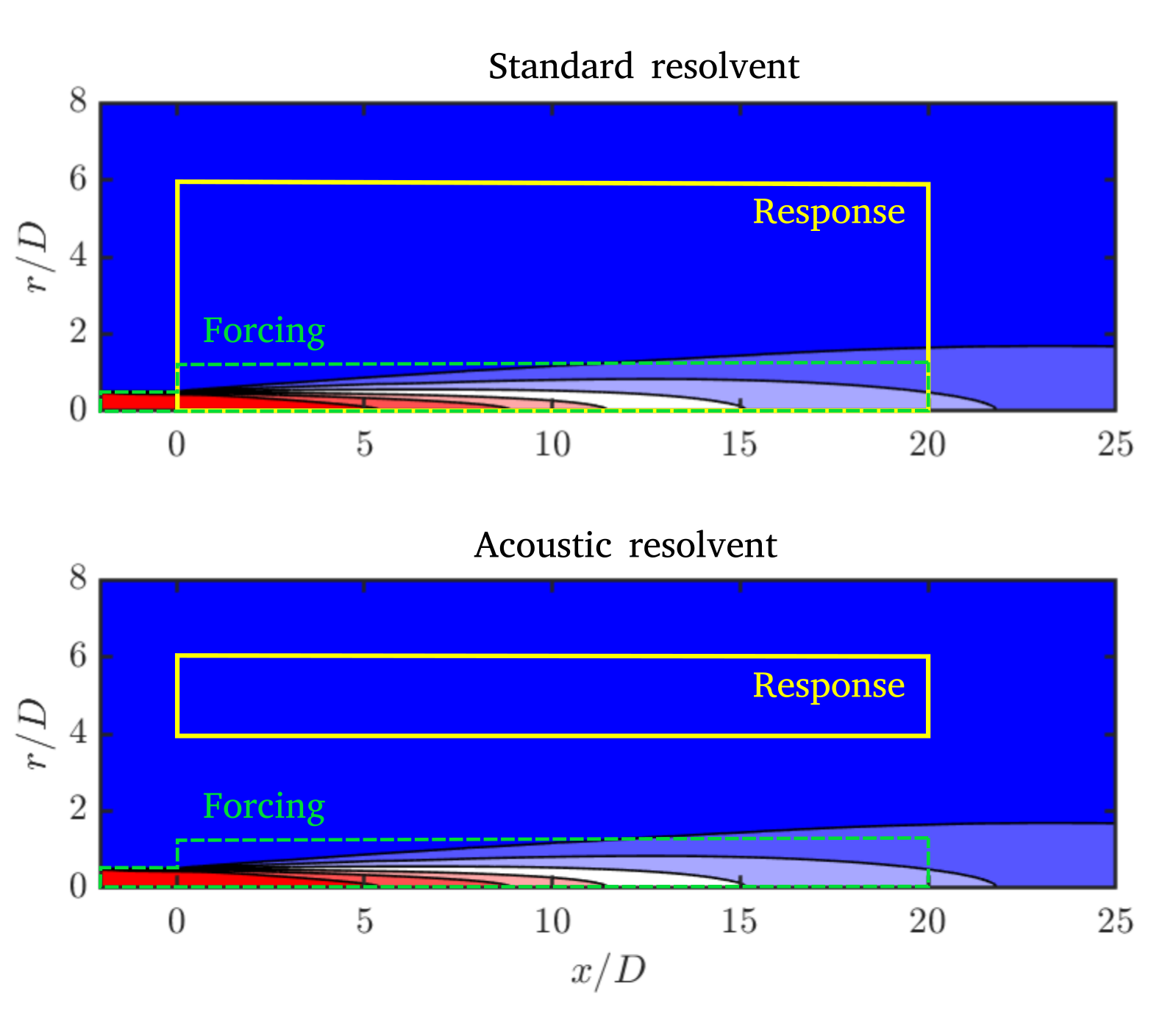}
    \caption{Locations of the forcing (green dashed line) and response (yellow solid line) optimisation regions in the standard and acoustic resolvents.
    Colour map is the same as in figure \ref{fig.meanflow}.} 
     \label{fig.regions}
\end{figure}

Two radically different choices of measurement regions of the response energy will be compared in this paper.
These choices are illustrated in figure \ref{fig.regions}.
As seen in section \ref{sec.norms}, the definition of this region is set through $\Omega$.
We will call \textit{standard} resolvent the approach in which this domain extends over $\rr/D  \in [0;6]$.
This includes the hydrodynamic field (i.e. mixing layer), which is usually the region of interest in resolvent analysis as it is where large energy growths takes place.
We will refer to \textit{acoustic} resolvent the calculations set with $\rr/D \in [4;6]$, which is located in the near acoustic field, away from the hydrodynamic region.
The remaining parameters are fixed as follows.
The optimisation region of the response axially extends over $\xr/D  \in [0;20]$.
The forcing region, set through the control matrix $\MatFlim$, extends over $\xf/D  \in [0;20]$ and $\rf/D  \in [0;1]$ in the jet region as well as in the nozzle over $\xf/D  \in [-2;0]$ and $\rf/D  \in [0;0.5]$. 

\subsection{Numerical method}

The resolvent code used to perform our calculation is based on that presented in \cite{bugeat20193d}.
This code is formulated from the non-linear compressible Navier-Stokes written in conservative form.
The resolvent matrix is constructed from the Jacobian matrix, which is obtained via a finite-difference approximation of the numerical residual of the non-linear equations, as proposed by \cite{mettot2014computation}.
Therefore, in this approach, the equations are first discretised and then linearised, bypassing the tedious task of linearising the continuous compressible equations before discretising them.
At this point, the Jacobian matrix is associated with conservative variables.
As we are interested, in particular, in pressure fluctuations in the acoustic field, singular vectors are expected to be expressed in terms of primitive variables.
When turbulent mean flows are considered, \cite{karban2020ambiguity} noted that a particular care is required in order to deduce the Jacobian matrix associated with a set of variables from that computed for another set.
The authors derived a transformation, which is used in the present paper in order to switch from conservative to primitive variables; details can be found in \cite{karban2020ambiguity}.

Following \cite{sipp2013characterization}, the SVD introduced in equation \eqref{eq.gsvd2} is solved as the following eigenvalue problem:
\begin{equation}
	\label{eq_RayleighQuotientEVP}		
		\underbrace{( \B{M}_F^{-1} \B{B}^* \B{R}^*  \B{M_R} \B{R} \B{B}) }_{\normalsize \B{A}} \B{v} = \sigma^2  \B{v} .
\end{equation}
The eigenvalue $\sigma^2$ and the eigenvector $\B{v}$ (which correspond to the singular value and right-singular vector of the aforementioned SVD) are solved with the the Krylov-Schur algorithm, using the SLEPC library \citep{Hernandez:2005:SSF}.
This requires the eduction of a Krylov subspace of the matrix $\B{A}$.
Each vector of this subspace is computed by sequentially solving two linear systems on the resolvent matrix, one involving $\B{R}$, the other involving $\B{R}^*$.
To this end, a direct LU method is employed using the PETSc library \citep{petsc-user-ref}.
This step constitutes the bottle neck of the algorithm in terms of both CPU time and RAM.
The complete procedure is detailed in \cite{bugeat20193d}.

The numerical domain extends over $x/D \in [-2.7;30]$ and $r/D \in [0;30]$.
This includes a nozzle flow in the region $x/D \in [-2.7;0]$ and $r/D \in [0;0.5]$.
We use Dirichlet boundary conditions at the inflow and at the top of the domain, and Neumann conditions at the downstream border.
Symmetry boundary conditions are used along the axis of the jet.
Along the nozzle wall, adiabatic, no slip conditions are set.

Sponge zones are implemented following \cite{agarwal2004calculation} and are located at $x/D>25$, $r/D>25$ and above the nozzle at $x/D<-0.3$.
Finally, a Cartesian mesh of size $595 \times 285$ is used throughout this paper. 
Grid independence of the results is shown in appendix \ref{ap.mesh}.



\section{Results} \label{sec.results}

\subsection{Comparison between standard and acoustic resolvents} \label{sec.standard_acoustic}

\begin{figure}
    \centering
    \includegraphics[angle=-0,trim=0 0 0 0, clip,width=1.00\textwidth]{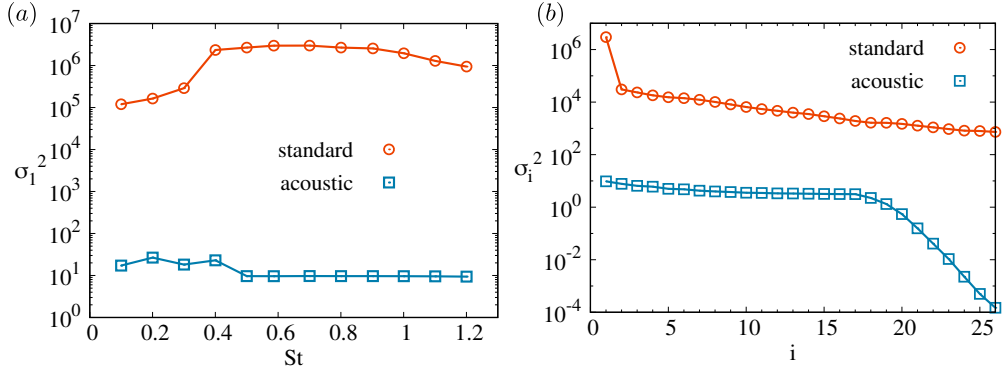}  
    \caption{Comparison between the SVD of the standard and acoustic resolvents. (a) First singular value at different frequency. (b) First twenty-six singular values at $St=0.6$.}
     \label{fig.SV_compareMethod}
\end{figure}

The first singular values of the standard resolvent are three to four orders of magnitude larger than those of the acoustic resolvent (figure \ref{fig.SV_compareMethod}a).
This is because the acoustic approach does not measure the large energy amplification in the mixing layer, which necessarily decreases the gain defined in equation \eqref{eq.gain}.
In the standard approach, higher frequencies ($St \gtrsim 0.4$) exhibit larger values, as also noted by several authors \citep{garnaud2013global,semeraro2016modeling,schmidt2018spectral}.
These authors showed that this corresponds to a switch of leading mode: Kelvin-Helmholtz (K-H) wavepackets here dominate the dynamics of the system whereas the Orr mechanism takes over at lower frequency.
In the acoustic approach, slightly larger singular values are observed at low frequencies.
They remain constant above $St=0.5$.

Now focusing on one frequency, $St=0.6$, the sub-optimal modes of the SVD are shown in figure \ref{fig.SV_compareMethod}(b).
In the standard approach, a well known gain separation of about two orders of magnitude is observed between $\sigma_1^2$ and $\sigma_2^2$.
This indicates that the single-point statistics of the system can be reasonably well modelled using only the first resolvent mode \citep{beneddine2016conditions}, which is associated with the K-H instability.
No such gain separation occurs for the acoustic resolvent. 
Instead, a slow decrease is observed for the first 18 singular values, after which a quick exponential decay takes place, as also noted by \cite{jeun2016input}.
The implication of this result regarding the construction of a low-rank response from resolvent modes is that several modes are likely to be required to build a response that recovers the energy in the selected acoustic domain.

\begin{figure}
    \centering
    \includegraphics[angle=-0,width=1.00\textwidth]{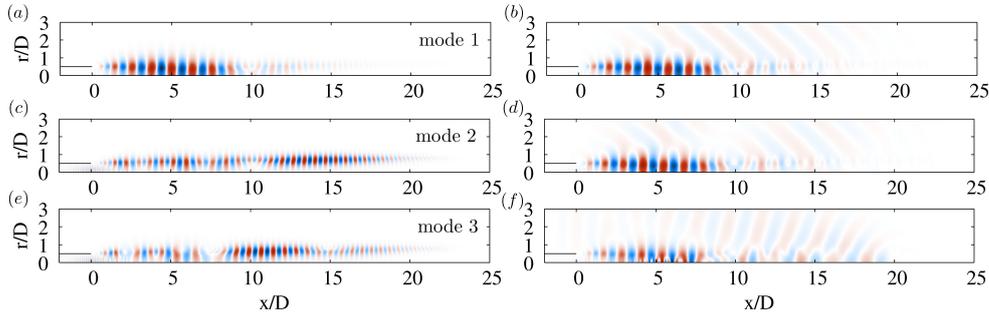} 
    \caption{Real part of the pressure of the three first response modes at $St=0.6$. 
    Left: standard resolvent. Right: acoustic resolvent.
    The horizontal black line represents the wall of the nozzle.}
     \label{fig.fields_p_compareMethod}
\end{figure}

\begin{figure}
    \centering    
    \includegraphics[angle=-0,trim=0 0 0 0, width=1.00\textwidth]{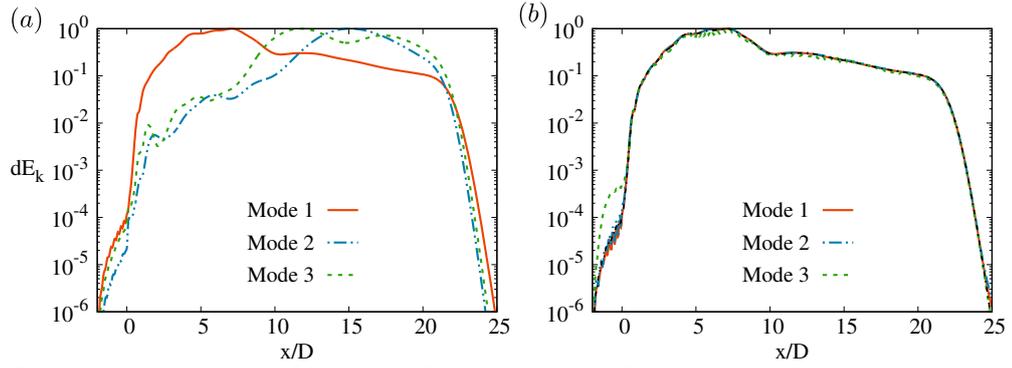}
    \caption{Kinetic energy density profiles of the three first resolvent response modes at $St=0.6$ for the standard (a) and acoustic (b) resolvents. 
    The black dotted line in (b) is the profile of mode 1 in (a). 
    Each profile is normalised by its maximum value.}
     \label{fig.profileEnergy_compareMethod}
\end{figure}

The hydrodynamic field of the first three response modes are shown for both approaches in figure \ref{fig.fields_p_compareMethod}.
{}In the standard approach, mode 1 is solely driven by the K-H wavepacket (figure \ref{fig.fields_p_compareMethod}a).
Sub-optimal modes differs from it, featuring structures developing further downstream with noticeably shorter wavelength, i.e. smaller phase velocity (figure \ref{fig.fields_p_compareMethod}c,e).
They are associated with the Orr mechanism \citep{lesshafft2019resolvent}.
As for the acoustic resolvent, a K-H wavepacket now appears in every mode (figure \ref{fig.fields_p_compareMethod}b,d,f).
To further examine these wavepackets, the kinetic energy density profiles along the axial direction, $\mathrm{d}E_k(x) = \int \frac{1}{2} \bar{\rho} |\mb{u} |^2 r \mathrm{d}r$, are calculated.
The K-H wavepacket in mode 1 grows exponentially immediately at the nozzle exit and reaches its maximum around $x/D \simeq 6$ (figure \ref{fig.profileEnergy_compareMethod}a).
Sub-optimal modes of the standard resolvent exhibit different behaviours as the maximum of each mode is reached at a different position from one another.
In the acoustic approach, however, energy profiles are the extremely similar for each mode outside of the nozzle (figure \ref{fig.profileEnergy_compareMethod}b).
Furthermore, they are nearly identical to that of mode 1 of the standard resolvent (K-H wavepacket).
The following interpretation can be proposed.
Sub-optimal response modes do not feature a K-H wavepacket in the standard approach because of the orthogonality constraint, which does not allow it.
Indeed, resolvent modes are constructed to be orthogonal to one another with respect to a scalar product defined from a choice of norm and a selected portion of the domain (section \ref{sec.norms}).
The standard approach includes the mixing layer zone in this domain, preventing the modes from sharing identical components.
In the acoustic approach, response modes do not have to be orthogonal in this region, and can therefore share the same behaviour.
The flow being linearly unstable, K-H wavepacket naturally appears if no orthogonality constraints aim at cancelling it.
As a result, it is rather expected to observe K-H wavepackets in all acoustic resolvent modes.




\begin{figure}
    \centering    
    \includegraphics[angle=-0,trim=0 0 0 0, clip,width=0.94\textwidth]{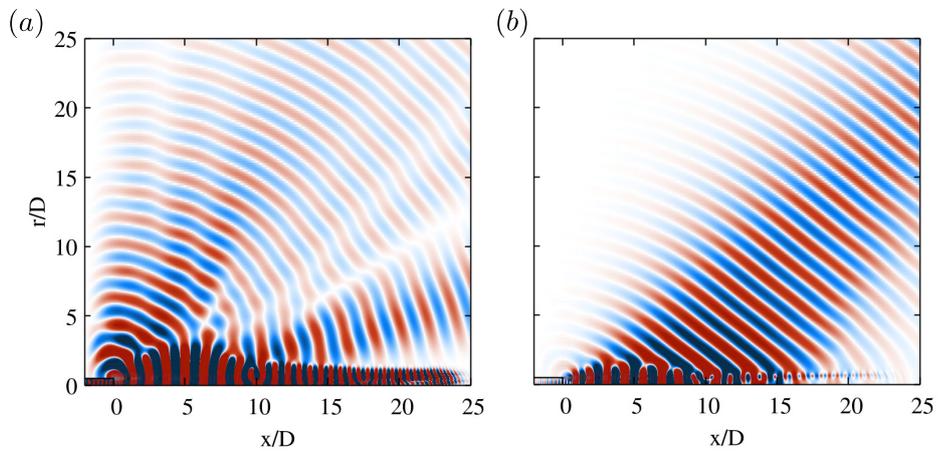}  
    \caption{Real part of the pressure of the optimal response mode at $St=0.6$ of the standard (a) and acoustic (b) resolvents. 
    In order to make the acoustic field visible, the colour bar is capped, in each case, by the maximum absolute value reached over $r/D>4$. 
    Note that these modes are the same as those already shown in figure \ref{fig.fields_p_compareMethod}.}
     \label{fig.fields_p_compareMethod_Acoustic}
\end{figure}

The acoustic field of the optimal mode is made visible in figure \ref{fig.fields_p_compareMethod_Acoustic}.
In the standard resolvent framework, a lot of the energy in the acoustic field results from effects associated with the nozzle, such as scattering or resonance.
In the acoustic approach, the acoustic field is more organised, featuring a clear directivity.
It is very similar to the optimal mode obtained by \cite{jeun2016input}.
It is striking to observe such different acoustic radiations while the wavepacket profiles are nearly identical (figure \ref{fig.profileEnergy_compareMethod}b).
This raises questions about the nature of the acoustic sources at play, which will be discussed in section \ref{sec.discussion} after having analysed more thoroughly the acoustic resolvent approach.


\subsection{Nozzle influence} \label{sec.nozzle}

\begin{figure}
    \centering
    \includegraphics[angle=-0,width=1.00\textwidth]{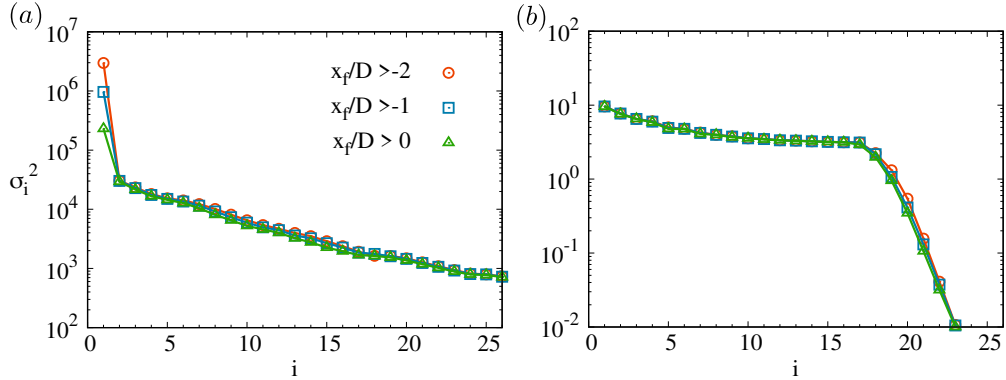}  
    \caption{Influence of the upstream location of the forcing region on the SVD at $St=0.6$. (a) Standard resolvent. (b) Acoustic resolvent.}
     \label{fig.SV_compareXf}
\end{figure}

The influence of the nozzle flow is investigated as it was not included in previous studies \citep{jeun2016input,schmidt2018spectral,pickering2021resolvent}.
Both standard and acoustic resolvent approaches are tested using different forcing regions selected through the control matrix~$\MatFlim$ (equation \eqref{eq.reduced_variables2}).
In the previous section, this region extended over $\xf/D \in [-2;20]$, recalling that the nozzle is located in the part of the domain where $x<0$.
We will now restrict the forcing domain to $\xf/D \in [-1;20]$ and $\xf/D \in [0;20]$, the latter meaning that strictly no forcing component is located inside the nozzle.
The influence of this parameter on the singular values is studied at $St=0.6$, which lies in the typical range of frequency of interest for jet noise.
In the standard approach, the first singular value increases as $\xf$ is decreased (figure \ref{fig.SV_compareXf}a).
The following singular values, however, remain unchanged.
This results in significantly larger gain separations as the forcing region is set more upstream in the nozzle.
This supports recent works that stressed the importance of the nozzle in the amplification of perturbations \citep{bres2018importance,lesshafft2019resolvent,kaplan2021nozzle}.
In contrast, the singular values of the acoustic resolvent remain unchanged by the variation of $\xf$ (figure \ref{fig.SV_compareXf}b).

\begin{figure}
    \centering
    \includegraphics[angle=-0,width=1.00\textwidth]{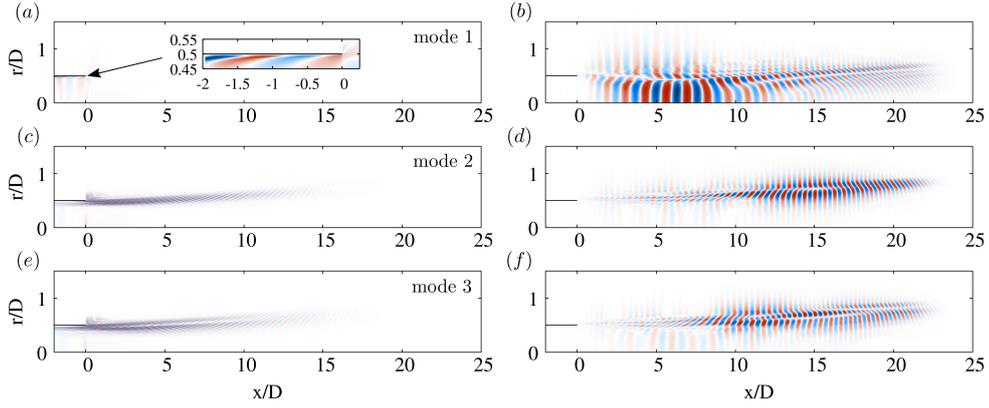} 
    \caption{Real part of the streamwise velocity of the three first resolvent modes at $St=0.6$. 
    Left: Forcing. Right: Response. 
    Results are for the standard resolvent only, using $\xf/D \in [-2;20]$.}
     \label{fig.fields_ux_Xfm2}
\end{figure}

\begin{figure}
    \centering
    \includegraphics[angle=-0,width=1.00\textwidth]{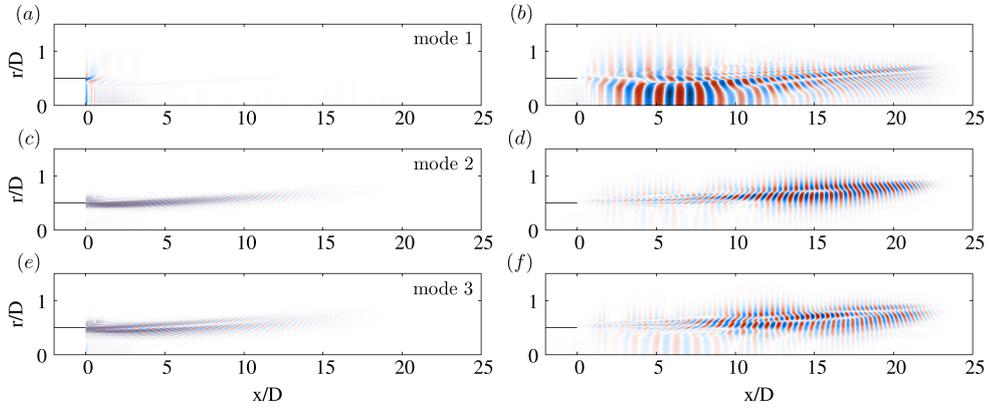} 
    \caption{Same as figure \ref{fig.fields_ux_Xfm2} but using $\xf/D \in [0;20]$.}
     \label{fig.fields_ux_Xf0}
\end{figure}

Examining the structure of the streamwise velocity fields of the standard resolvent modes can shed light on the behaviour of their associated singular value (figure \ref{fig.fields_ux_Xfm2} and \ref{fig.fields_ux_Xf0}).
It is not surprising that both the forcing and response of the sub-optimal modes for $\xf/D>-2$ and $\xf/D>0$ are identical given that their singular values are the same.
Interestingly, the optimal response is the same whether the forcing field is allowed in the nozzle or not (figure \ref{fig.fields_ux_Xfm2}b and \ref{fig.fields_ux_Xf0}b).
Thus, the difference found for the first singular value must be explained by a change in the forcing mode.
In the $\xf/D>0$ case, the forcing is concentrated in the mixing layer at the immediate exit of the nozzle (figure \ref{fig.fields_ux_Xf0}a).
In the $\xf/D>-2$ case, it is spread in the near-wall region of the nozzle (figure \ref{fig.fields_ux_Xfm2}a). 
Both forcing fields tend to be located as upstream as possible.
This is indeed the most efficient way to promote energy amplification as the growth of convective instabilities is then experienced over longer distances.
Besides, the tilted forcing structure spreading in the nozzle shows the action of the Orr mechanism in the boundary layer.
This non-modal growth occurs all along the nozzle length, further supporting that a longer nozzle induces larger energy growth.
The sub-optimal forcing modes also feature tilted structures but are mostly located in the jet region rather than in the nozzle, which explains why the sub-optimal singular values are insensitive to $\xf$.
These modes are purely driven by the Orr mechanism, which is at play in the mixing layer, as analysed by \cite{lesshafft2019resolvent}.



\subsection{Focus on the acoustic resolvent} \label{sec.acousticResolvent}

The work presented so far has dealt with highlighting some fundamental differences between the standard and acoustic resolvents.
From now on, we will focus mostly on the acoustic resolvent.
The influence of the forcing region will first be studied.
Afterwards, a detailed analysis of the optimal and sub-optimal modes will be carried out before assessing the potential of the acoustic resolvent to build a low-rank model of jet noise.
The effect of the Mach number will eventually be examined.

\subsubsection{Sensitivity to the radial extent of the forcing region} \label{sec.forcing_region}

\begin{figure}
    \centering
    \includegraphics[angle=-0,trim=0 0 0 0,width=1.00\textwidth]{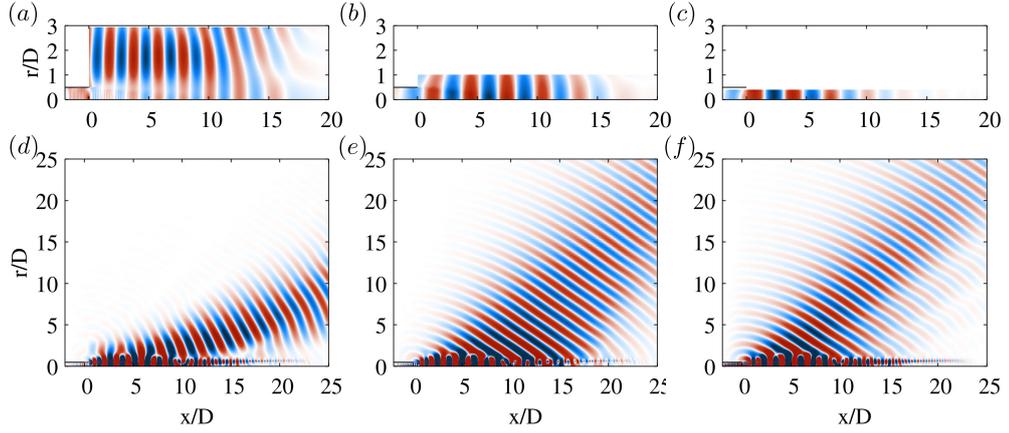}    
    \caption{Real part of the pressure of the first resolvent mode at $St=0.6$. 
    Top: forcing. Bottom: response. 
    Left: F3 case. Centre: reference case. Right: F04 case. 
    The nomenclature is given in table \ref{tab.casesForcing}. 
    Colour bars of the response modes are capped as in figure~\ref{fig.fields_p_compareMethod_Acoustic}.}
     \label{fig.fieldL25_mode1_compareRf}
\end{figure}

Three forcing regions of different radial extent are considered (see table \ref{tab.casesForcing}).
The optimal forcing and response modes for each case are presented in figure \ref{fig.fieldL25_mode1_compareRf}.
Each forcing mode features a streamwise wave structure. 
In the F3 case, most of the forcing is located above the mixing layer and has a noticeably shorter wavelength than in the other cases (figure \ref{fig.fieldL25_mode1_compareRf}a).
The associated response is made of a low-angle acoustic beam (figure \ref{fig.fieldL25_mode1_compareRf}d).
Without being strictly identical, the reference and F04 cases are similar to each other, but greatly differ from the F3 set-up.
Their optimal response features an acoustic radiation of larger angle (figure \ref{fig.fieldL25_mode1_compareRf}e,f).
The reference and F04 cases constitute a more realistic constraint on the forcing (figure \ref{fig.fieldL25_mode1_compareRf}b,c) compared to the F3 case, given that the forcing is actually limited within the shear layer as shown by \cite{towne2017statistical}. 
The region $r\in[1,3]$ hardly contains any turbulent activity, and hence, forcing. 
From a hydrodynamic point of view, the response mode includes a K-H wavepacket in each case -- even in the case $\rf/D \in [0;0.4],$ which has a zero-forcing close to the wall of the nozzle and at the start of the mixing layer.
These regions were found to be the most efficient locations to excite this wavepacket (section \ref{sec.nozzle}).
However, it is not a necessary condition to force the system in these regions in order to generate this wavepacket; as mentioned in section \ref{sec.standard_acoustic}, in the absence of orthogonality constraints in the mixing layer, these wavepacket will grow regardless of the shape and the precise location of the upstream forcing as they result from a linear instability of the flow.

\begin{table}
 \begin{tabular}{lllc}
     \toprule
     Cases \hspace{0.52cm} & Forcing pipe \hspace{0.8cm}  & Forcing jet \hspace{0.6cm}  &   Response \hspace{0.2cm}\\ 
     \midrule
     Acoustic resolvent F3         & $[-2;0] \times [0;0.5]$   & $[0;20] \times [0;3]$ &  $[0;20] \times [4;6]$    \\
     Acoustic resolvent Reference  & $[-2;0] \times [0;0.5]$   & $[0;20] \times [0;1]$ &  $[0;20] \times [4;6]$    \\
     Acoustic resolvent F04        & $[-2;0] \times [0;0.4]$   & $[0;20] \times [0;0.4]$ &  $[0;20] \times [4;6]$  \\
     \botrule
 \end{tabular}
 \caption{Nomenclature of the different cases studied in section \ref{sec.forcing_region}. }
 \label{tab.casesForcing}
\end{table}

\subsubsection{Physical features of the acoustic resolvent modes} \label{sec.subopt}

\begin{figure}
    \centering
    \includegraphics[angle=-0,trim=0 0 0 0,width=1.00\textwidth]{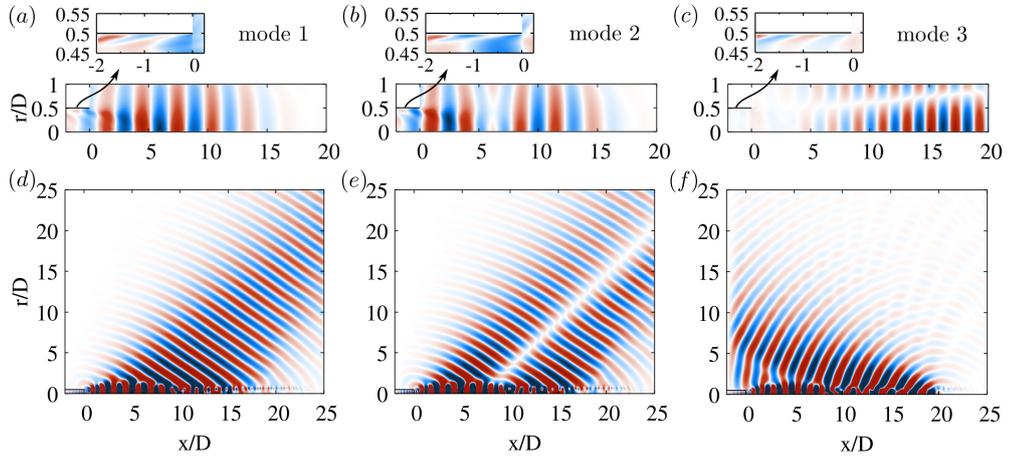}    
    \caption{First three modes of the acoustic resolvent (reference case) at $St=0.6$. 
    Top: forcing, real part of the streamwise velocity. Bottom: response, real part of the pressure is shown. 
    Colour bars of the response modes are capped as in figure \ref{fig.fields_p_compareMethod_Acoustic}.}
    \label{fig.field_p_subopt}
\end{figure}

In this section, we focus on the reference case of the acoustic resolvent and discuss the first three resolvent modes.
Note that hydrodynamic fields have already been shown in figure \ref{fig.fields_p_compareMethod}, where K-H wavepackets were observed in each response mode.
Unlike mode 1 (figure \ref{fig.field_p_subopt}d), the acoustic radiation in mode 2 is now made of two distinct beams (figure \ref{fig.field_p_subopt}e).
This behaviour was also reported by \cite{jeun2016input}.
This results from the orthogonality constraint in the acoustic field. 
The forcing field of mode 2 follows the same trend (figure \ref{fig.field_p_subopt}b): the streamwise wave that is found in mode 1 (figure \ref{fig.field_p_subopt}a) is now divided into two portions of space.
Mode 3 is very different as the spatial support of the forcing is essentially located downstream (figure \ref{fig.field_p_subopt}c).
Moreover, the wavelength appears shorter.
In fact, we will show that the phase velocity is negative: the forcing field is essentially an upstream-travelling wave.
As a result, the associated acoustic radiation is made of an upstream propagating wave with a clear directivity and emitted from $x/D \simeq 20$ (figure \ref{fig.field_p_subopt}f).
In addition, each forcing mode contains a tilted structure near the nozzle, similar to the structure that was found to generate the optimal hydrodynamic response in the standard resolvent (figure \ref{fig.fields_ux_Xfm2}a). 
Even though it is not dominant -- most of the energy of the forcing is located downstream of the nozzle, this indicates that K-H wavepackets play a role in optimal sound-generation.

\begin{figure}
    \centering    
    \includegraphics[angle=-0,trim=0 0 0 0, clip,width=1.00\textwidth]{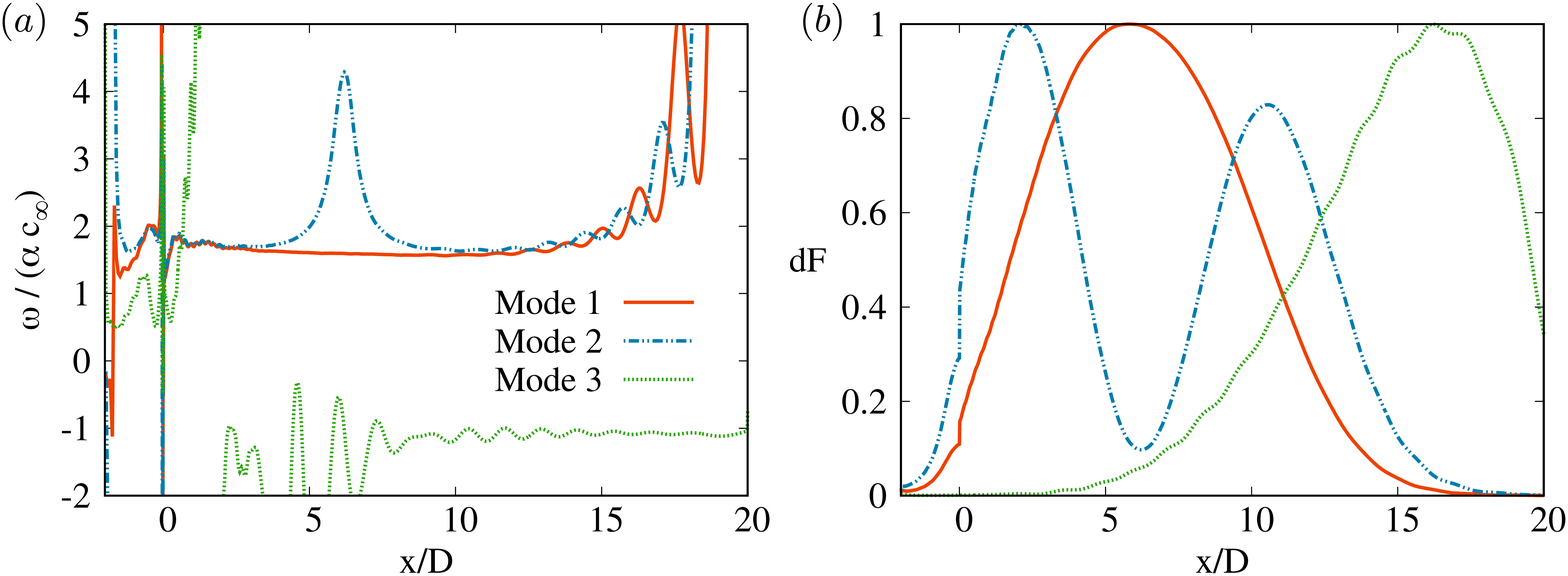}
    \caption{Phase velocity along $r/D=0.49$ (a) and energy density profile (b) of the first three forcing modes of the acoustic resolvent (reference case) at $St=0.6$.} 
     \label{fig.profile_phaseVel_Energy_Forcing}
\end{figure}

The phase velocity of the forcing field can be computed as $\cphi = \omega / \alpha$ if we assume it is a plane wave in the streamwise direction.
The wavenumber can then be approximated by  $\alpha = \derp{\theta}{x}$ where $\theta$ is the phase of the complex pressure field.
Phase velocity profiles of the first three forcing modes are shown in figure \ref{fig.profile_phaseVel_Energy_Forcing}(a).
Let us first note that diverging phase values and spurious oscillations are observed in some portion of the profiles. 
This is either caused by additional waves associated with nozzle effects or because the plane wave assumption does not hold any longer. 
However, clean phase velocities are obtained in regions of large forcing, which allows us to comment on the prevalent forcing mechanisms.
Large forcing locations can indeed be detected by plotting the forcing energy density $\mathrm{d}F(x)$ (figure \ref{fig.profile_phaseVel_Energy_Forcing}b), which is defined analogously to the kinetic energy density in section \ref{sec.standard_acoustic}.
Forcing mode 1 and 2 have a supersonic phase velocity with $\cphi/a_\infty \simeq 1.6$ at the location of maximum energy density, where $a_\infty$ is the speed of sound in the far field.
Supersonic waves are promoted by the acoustic resolvent calculation because they are efficient acoustic sources \citep{crighton1975basic}, maximising the acoustic energy for a given energy input.
Overall, three waves of different phase velocities coexist: the supersonic forcing, the K-H wavepacket and the acoustic radiation.
Forcing mode 3 is also slightly supersonic and has a negative phase velocity.
Its energy clearly peaks in the downstream region, where an upstream-travelling acoustic wave is radiated.
Such modes are unphysical in jet noise, as will be discussed later.

The direct link between the forcing and acoustic fields can be clearly shown by relating the angle of the radiation $\theta$ to the phase velocity of the forcing which, for a Mach wave, verifies $\theta = \cos^{-1} \left( \frac{a_\infty}{c_\varphi} \right)$.
The angle is defined with respect to the centre axis and a value of $0^\circ$ would correspond to downstream radiation.
From the calculation of the phase velocity at the location of the maximum forcing field, the predicted angle is $\theta = 51.3^\circ$.  
Meanwhile, the acoustic radiation observed in the response mode is $\theta \simeq 51^\circ$.    
While this confirms that the forcing field is directly responsible for the acoustic radiation, it is unclear why this specific angle $\theta$ is selected by the resolvent analysis.
Indeed, the phase velocity of the forcing field could \textit{a priori} take any values: introduced as an external body force, it is not constrained by the hydrodynamics.
The comparison with SPOD modes carried out in the next section will shed more light on this matter.

\subsubsection{Comparison with SPOD modes} \label{sec.spod}

\begin{figure}
    \centering
    \includegraphics[angle=-90,trim=0 0 0 0, clip,width=0.5\textwidth]{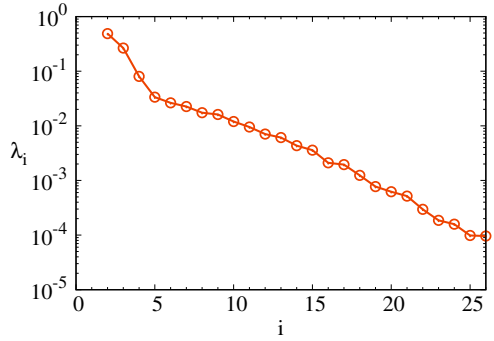} 
    \caption{Eigenvalues of the acoustic SPOD at $St=0.6$.}
     \label{fig.SPODeigenvalues}
\end{figure}

Calculated from the times series obtained from LES calculations, SPOD modes provide, for each frequency, an optimal orthogonal basis to represent the acoustic field.
These modes are obtained through the eigendecomposition of the cross-spectral density (CSD) matrix. 
\cite{towne2018spectral} gave a comprehensive presentation of their calculation and properties.
SPOD modes are associated with a norm.
The same set-up as that of the resolvent modes is used: Chu's energy is calculated over the acoustic domain of the response. 
The normalised eigenvalues of the SPOD modes at $St=0.6$ are shown in figure \ref{fig.SPODeigenvalues}, indicating the proportion of the total energy that each mode contains.
Contrary to the SVD of the acoustic resolvent (figure \ref{fig.SV_compareMethod}a), the SPOD is low-rank.
A separation of more than one order of magnitude is observed between the first and fourth SPOD modes while it took nineteen resolvent modes to achieve such a separation.
This means that the acoustic field can be described by a low-rank basis, but that the current resolvent approach does not provide this basis -- at least without further information.

\begin{figure}
    \centering    
	\includegraphics[angle=-0,trim=0 0 0 0, clip,width=1.00\textwidth]{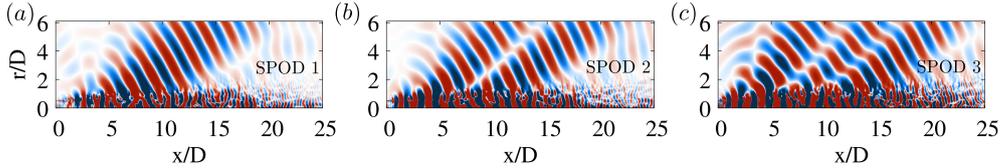} 
    \caption{Three first SPOD modes at $St=0.6$, real part of the pressure.
    Colour bars are capped as in figure \ref{fig.fields_p_compareMethod_Acoustic}.}
    \label{fig.field_p_SPOD}
\end{figure}

However, the acoustic resolvent contains relevant features that were not present in the standard resolvent approach.
This can be seen by examining the first three SPOD modes (figure \ref{fig.field_p_SPOD}).
They are made of a clear acoustic radiation which, because of the orthogonality constraint, is divided into different beams as the rank of the mode increases.
This feature was also found in the two first modes of the acoustic resolvent, but absent in the standard resolvent decomposition.
Moreover, the angle of radiation of the first mode is about $54^\circ$, close to the value of $51^\circ$ found for the Mach wave in the first resolvent mode.
This indicates that the phase velocity of the forcing at the origin of this Mach wave must somehow be rooted in the mean flow properties. 
The structure of the third SPOD mode confirms that the third mode of the acoustic resolvent modes, made of an upstream propagating acoustic wave (figure \ref{fig.field_p_subopt}f), is indeed not relevant to the acoustics -- even though this mode is optimal in the given resolvent framework, in which any external forcing is allowed.


\begin{figure}
    \centering
    \includegraphics[angle=-90,trim=0 0 0 0, clip,width=0.5\textwidth]{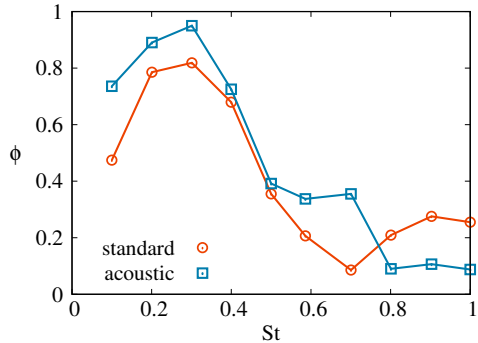} 
    \caption{Alignment between SPOD and resolvent modes.}
     \label{fig.alignmentSPOD}
\end{figure}

\begin{figure}
    \centering
    \includegraphics[angle=-0,trim=0 0 0 0, clip,width=0.995\textwidth]{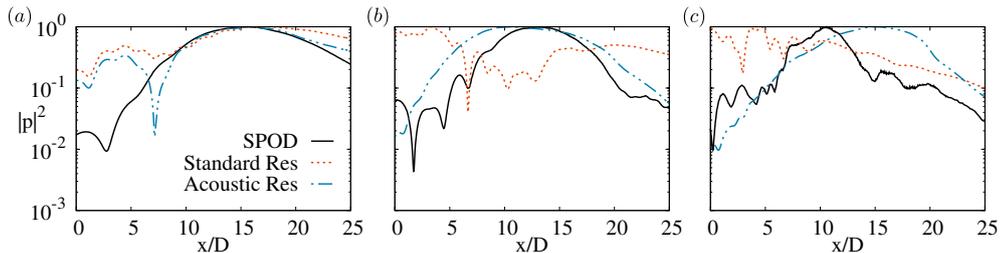} 
    \caption{Pressure along the line $r/D = 5$ in the near acoustic field at $St=0.3$ (a), $St=0.6$ (b) and $St=1$ (c).}
    \label{fig.acousticLine}
\end{figure}

The alignment between SPOD and resolvent modes can be assessed through a more global quantity, $\phi$, defined as a normalised scalar product between these modes:
\begin{equation}
    \phi = \frac{\langle q_\mathrm{res} , q_\mathrm{SPOD} \rangle}{\sqrt{ \langle q_\mathrm{res} , q_\mathrm{res}\rangle \langle q_\mathrm{SPOD} , q_\mathrm{SPOD} \rangle } } .
\end{equation}
The scalar product is associated with Chu's energy taken over the near-acoustic domain $x \in [0;25]$ and $r \in [4;6]$.
The first SPOD and resolvent modes are considered.
The results are presented for different frequencies in figure \ref{fig.alignmentSPOD}.
The acoustic resolvent generally produces better alignment with SPOD modes than the standard resolvent, except at the highest frequencies considered ($St \simeq 1$).
The differences between the two approaches are not dramatically large but remain significant.
Both exhibit a maximum of alignment around $St=0.3$ where the acoustic resolvent reaches a value of $\phi$ close to 1.
A steep decrease is observed above $St > 0.4$.
There, values of $\phi$ do not drop to zero but rather stay around 0.1 and 0.4, indicating that some acoustic features may still be captured by the resolvent decomposition.
In order to better apprehend the correspondence between these values of $\phi$ and the near-acoustic field, the pressure is plotted along the line $r/D=5$ for each mode at three different frequencies (figure~\ref{fig.acousticLine}). 
Resolvent and SPOD modes are compared, and the amplitude of each mode is normalised by its own maximum value.
As expected, the maximum of acoustic radiation is particularly well predicted at $St=0.3$ by the acoustic resolvent (figure~\ref{fig.acousticLine}a).
The pressure profile is particularly well recovered between around $x/D=10$ and $x/D=20$, where the signal is the largest.
The standard resolvent also performs well but is not as accurate, predicting a peak further downstream than that of the SPOD mode, and a decay with $x$ that is not as fast as in the SPOD mode.
The standard resolvent at $St=0.6$, where $\phi \simeq 0.2$, predicts none of the behaviour of the SPOD mode (figure~\ref{fig.acousticLine}b).
Conversely, with a better alignment at this frequency ($\phi \simeq 0.35$), the acoustic resolvent mode is able to detect the peak of pressure of the SPOD mode but does not correctly capture its shape.
This tempers the above discussion about the angle of the acoustic beams observed in the SPOD and resolvent response fields.
Finally, at $St=1$, both approaches produce poor predictions (figure~\ref{fig.acousticLine}c).


\subsubsection{Effect of Mach number}

\begin{figure}
    \centering
    \includegraphics[angle=-0,trim=0 0 0 0, clip,width=0.97\textwidth]{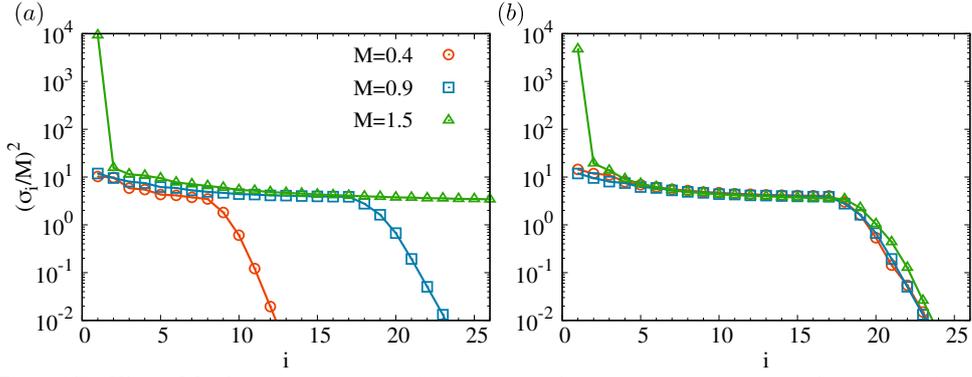} 
    \caption{First 26 singular values of the acoustic resolvent, renormalised by $M$, at $St=0.6$ (a) and $St_a= St \times M = 0.53$ (b) for three Mach numbers.}
    \label{fig.SV_compareMach}
\end{figure}

\begin{figure}
    \centering
    \includegraphics[angle=-0,trim=0 0 0 0, clip,width=0.995\textwidth]{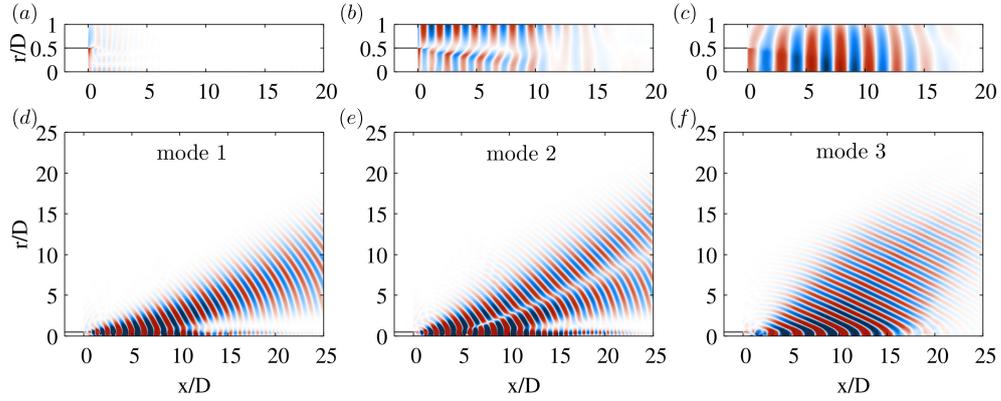} 
    \caption{Same as figure \ref{fig.field_p_subopt} but at $M=1.5$ ($St=0.6$). Note that the numerical domain in the supersonic case only starts at $x/D=0$, but a larger domain is displayed to ease comparison with other cases.}
     \label{fig.field_p_suboptM1_5}
\end{figure}

So far, only the case $M=0.9$ has been studied.
Two additional Mach numbers are now considered, $M=0.4$ and $M=1.5$, whose mean flows have also been obtained by LES calculations.
The comparison of the singular values of the acoustic resolvent at $St=0.6$ for the three Mach numbers is shown in figure \ref{fig.SV_compareMach}(a).
The supersonic case has a distinct behaviour as a separation of nearly three orders of magnitude is found between $\sigma_1^2$ and $\sigma_2^2$, as also noted by \cite{jeun2016input}.
This results from the existence of a supersonic instability wave which directly generates a Mach wave: the optimal forcing is located at the most upstream location (figure \ref{fig.field_p_suboptM1_5}a) in order to trigger a K-H wavepacket to which is attached the acoustic radiation (figure \ref{fig.field_p_suboptM1_5}d).
In other words, the most efficient way to generate acoustic perturbations coincides with the most efficient way to produce hydrodynamic perturbations.
This is different from the subsonic case previously investigated, in which the Mach wave was dominantly produced by the supersonic forcing field rather than the response field.
Note that the available data from LES calculation in the supersonic case does not contain a nozzle: the mean flow used for the resolvent calculation starts at $x/D=0$.
If it was included, it is very likely that the optimal forcing would be located upstream in the nozzle as in the $M=0.9$ case of the standard resolvent (figure \ref{fig.fields_ux_Xfm2}a), and that the gain separation would be even larger.
The second resolvent mode of the $M=1.5$ case features two acoustic beams with a similar angle to that of the first mode (figure \ref{fig.field_p_suboptM1_5}e).
It does not result from an instability wave, but is rather directly produced by the forcing as in the subsonic case.
If the instability wave was triggered, it would indeed generate the same acoustic radiation as that of the first mode, hence breaking the orthogonality constraint.
Therefore, the sub-optimal modes cannot contain the supersonic K-H wavepacket.
Interestingly, the third mode does not divide the acoustic field into a third beam (figure \ref{fig.field_p_suboptM1_5}f), but rather contains a single radiation of different angle than the two other mode.
However, the structures of sub-optimal modes do not crucially matter in the supersonic regime since, given the large gain separation, the K-H wavepacket associated with first resolvent mode is sufficient to model jet noise in this case \citep{sinha2014wavepacket}.


\begin{figure}
    \centering
    \includegraphics[angle=-0,trim=0 0 0 0, clip,width=0.995\textwidth]{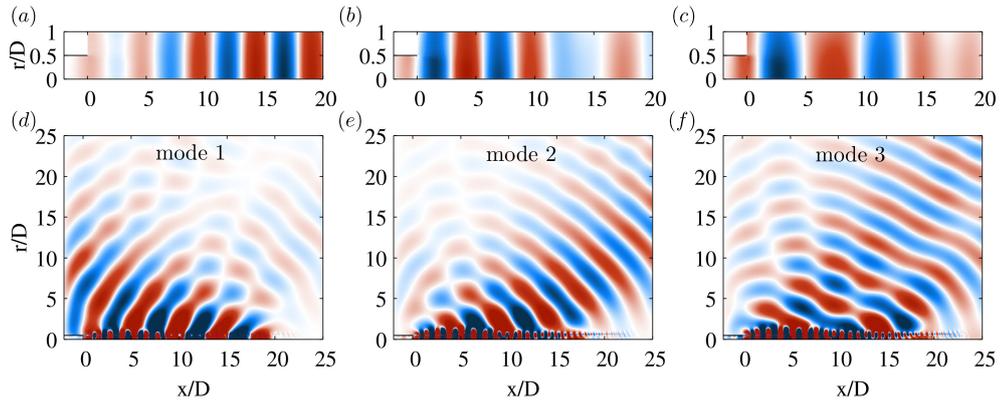} 
    \caption{Same as figure \ref{fig.field_p_subopt} but at $M=0.4$ ($St=0.6$)}
     \label{fig.field_p_suboptM04}
\end{figure}

The behaviour of the singular values at $M=0.4$ case is similar to that of the $M=0.9$ case (figure \ref{fig.SV_compareMach}a).
The first singular values have a comparable magnitude before experiencing a quick drop.
This drop occurs at a smaller rank than that observed for $M=0.9$.
Once again, the forcing modes contain supersonic waves generating Mach waves (figure \ref{fig.field_p_suboptM04}a,b,c).
The first mode is seen to be upstream propagating which, again, indicates the inability of the acoustic resolvent to produce a hierarchy of relevant acoustic modes.

\begin{figure}
    \centering
    \includegraphics[angle=-0,trim=0 0 0 0, clip,width=0.995\textwidth]{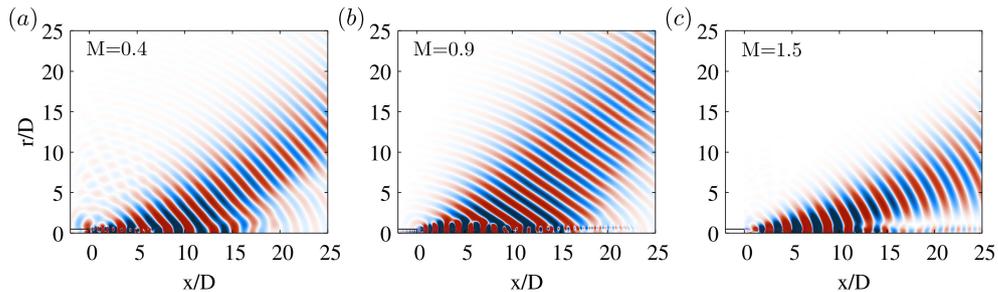} 
    \caption{Optimal response mode of the acoustic resolvent at three Mach numbers at constant Helmholtz $St_a=0.53$.
    Colour bars of the response modes are capped as in figure \ref{fig.fields_p_compareMethod_Acoustic}.}
     \label{fig.field_p_compareMach}
\end{figure}

Coming back to the singular values presented in figure \ref{fig.SV_compareMach}(a), it can be seen that a normalisation of $\sigma_i$ by $M$ was employed.
By doing so, the reference velocity scale is no longer that at the nozzle exit, but rather the speed of sound in the quiet flow.
This is Helmholtz scaling.
This choice of normalisation produces singular values of similar magnitudes for any Mach numbers.
This indicates that these singular values are associated with the same acoustic mechanism, seen to be linked to the Mach-wave mechanism.
The exception is the previously discussed first singular value at $M=1.5$, associated instead with a K-H wavepacket, and whose magnitude is indeed much larger. 
To examine this further, it is then interesting to compare the behaviour of the different Mach numbers at a constant Helmholtz number, defined as $St_a = St \times M$.
In this case, at $St_a = 0.53$, the singular values collapse for any Mach numbers (figure~\ref{fig.SV_compareMach}b), an observation first made by \cite{jeun2016input}.
Therefore, the speed of sound is the relevant velocity scale of the optimisation problem, consistent with the importance of acoustic matching central to the jet-noise problem \citep{crighton1975basic}.
The optimal response modes at this constant Helmholtz number $St_a$ are presented in figure \ref{fig.field_p_compareMach}.
All modes now have the same acoustic wave length ($\lambda / D= 1/St_a$).
At this frequency, all optimal modes represent downstream-travelling waves. 
In the supersonic case, the acoustic radiation is clearly linked to the hydrodynamic wavepacket, both fields being synchronised.
In the subsonic cases, the acoustic wave length is distinct from that of the wavepacket.



\section{Discussion} \label{sec.discussion}

\subsection{Wavepackets}

In the subsonic regime, the SVD of the acoustic resolvent identifies planar supersonic waves as the optimal forcing mode, i.e. the most efficient way to generate energy in the near acoustic field.
While the forcing field varies for each mode, the hydrodynamic field observed in the response remains nearly identical for each of them (figure \ref{fig.profileEnergy_compareMethod}b), resulting from the linear convective instability of the flow.
In this picture, the role of hydrodynamic wavepackets, known to be central in jet noise \citep{jordan2013wave}, becomes unclear as the hydrodynamic properties of the flow appear decorrelated from the acoustic field.
We here propose an interpretation of these observations.

Subsonic jet noise is primarily understood as a result of acoustic matching between supersonic wavenumbers associated with the spatial envelop of subsonic wavepackets and the acoustic field \citep{crighton1975basic}.
These supersonic components represent a small fraction of the total energy of the hydrodynamic field: subsonic jet noise is inefficient in the sense that the ratio between the acoustic and hydrodynamic energies is very small (in contrast, supersonic instability waves found in supersonic jets are much more efficient).
The supersonic forcing waves promoted by the acoustic resolvent thus appear consistent with acoustic matching, as it pinpoints the essential mechanism of jet noise.
The fact that wavepackets do not appear as an input (forcing) of our analysis, but rather as an output (response) along with acoustic radiation, is associated with the reorganisation of the Navier-Stokes equations in the resolvent framework.
Here, all non-linear terms are seen as an input while the associated output develops according to linear mechanisms.
Wavepackets, which result from a linear amplification, are thus observed in the output modes.
Furthermore, in the resolvent framework, forcing modes are introduced as an external force acting on the flow; the SVD produces modes that are not constrained by the intrinsic dynamics of the flow and may thus take any shape, only aiming at satisfying the optimality constraint that is imposed.
These forcing modes are ultimately interpreted as a basis for the intrinsic non-linear turbulent interactions in the flow \citep{mckeon2010critical,hwang2010amplification}).
Therefore, the supersonic forcing structures can emerge from wavepackets that jitter on account of non-linear effects and contain associated supersonic wavenumbers allowing for acoustic matching.

\subsection{Perspectives for resolvent-based modelling of subsonic jet-noise}

A central message to be taken from our analysis is that, when performing resolvent analysis, the SVD alone should be considered with caution. While in systems exhibiting a clear gain separation leading modes can give insight into underlying mechanisms, in the absence of such gain separation, the resolvent operator and bases must be considered with circumspection. The jet-noise problem considered is one such case. And the non-physical, upstream-travelling modes we identify provide an eloquent illustration of this point. 

For problems in which gain separation is lacking---and we here have in mind jet noise---alternative strategies are required. Two examples are studies of \cite{pickering2021resolvent} and \citep{karban2022empirical}. 
In the former, the characteristics of the forcing projection into the input space of an acoustic resolvent operator were inferred from the sound field, and a low-rank model thus obtained. 
Including eddy viscosity and restricting the portion of the acoustic field to lower angles (which may prevent the existence of upstream travelling modes), a rank-1 model was proposed. 

A conclusion of the work we report here is that, for jet noise, forcing data is necessary to provide clarification of underlying mechanisms. Such data is, however, extremely difficult to obtain from experiment or simulation when the problem considered involves high Reynolds number, which is the case of interest for jet noise \citep[cf. ][]{karban2022solutions}. And when such data is available, interpreting and modelling the coupling between non-linear scale interaction and sound remains an daunting task. 

Having done the work necessary to obtain useable forcing data from a moderately resolved, high-Re LES in \cite{karban2022solutions}, a second effort \citep{karban2022empirical} was undertaken to understand how that data might be used to probe the mechanisms by which non-linear scale interactions drive jet noise. The difficulty of the problem faced can be appreciated by considering that less than 0.1$\%$ of the forcing fluctuation energy is responsible for the radiated sound. In \cite{karban2022empirical}, dedicated data-reduction was necessary to first isolate the forcing subspace correlated with the farfield sound, using the RESPOD technique developed in \cite{karban_jfm_2022}. But as this subspace is dominated by forcing components that are not acoustically matched, it provides neither clear physical interpretation nor guidance for simplified modelling.

\begin{figure}
    \centering
    \includegraphics[angle=-0,trim=0 0 0 0, clip,width=0.995\textwidth]{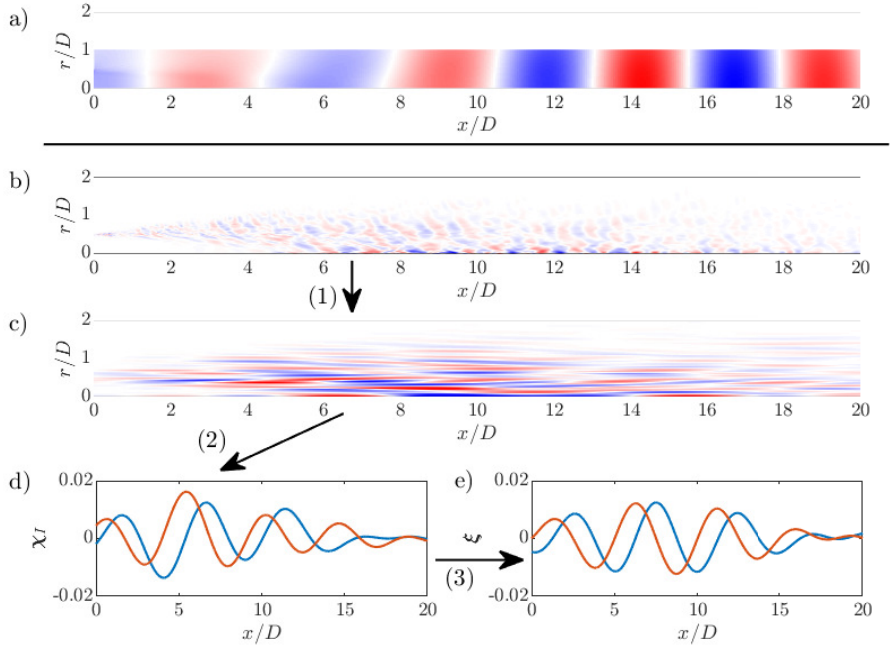} 
    \caption{Diagram showing the modelling steps pursued to obtain the empirical jet noise model introduced in \cite{karban2022empirical}. Given the constant supersonic wave structure with constant radial support observed in the optimal acoustic forcing mode (a), the RESPOD forcing mode (b), which generates $\sim80$\% of the total downstream acoustic energy, is (1) filtered to retain the components with supersonic phase speed (c), (2) integrated over the radial direction to obtain a line source (d), which is finally (3) modelled using an empirical model (e).} 
     \label{fig.modellingsteps}
\end{figure}

Those goals where achieved thanks to the results of the current paper: both the acoustic resolvent operator \textit{and} the input bases were crucial. We provide a brief explanation in what follows.

Despite the lack of a gain separation in the acoustic resolvent, the shapes of the modes provide the key to understanding the `acoustically matched' components of the forcing field: they show how the forcing field filters through the resolvent operator, and how that filtering operation extracts the acoustically matched components from the predominantly silent forcing field. The acoustic-resolvent input modes exhibit two salient traits: supersonic phase speed; almost no radial oscillation (cf. Figure \ref{fig.modellingsteps}-a). These traits motivated the final processing of the LES forcing data (Figure \ref{fig.modellingsteps}-(b-d)) and led to the elaboration of a rank-1 empirical source model (Figure \ref{fig.modellingsteps}-e). Without the acoustic resolvent operator we have developed and analysis of its SVD, the model of \cite{karban2022empirical} would not have been obtained. 

And it is noteworthy that the source model identified in that work is (thanks to the results of the present paper) characterised by a robustness that permits the computation of jet-noise over a range of operating conditions, including cases with jets in forward flight. Indeed it was possible to use the model, coupled with the acoustic resolvent reported here, in an industry environment. Airbus computed mean fields using industry-standard RANS. They used the code presented in the current paper to compute acoustic resolvent operators and then performed jet-noise calculations using the source model of the companion study \cite{karban2022empirical}. The accuracy of the downstream acoustic predictions across the range of operating conditions considered was within 2-3 dB.


\section{Conclusion} \label{sec.conclusion}

We have carried out an acoustic-oriented resolvent analysis on a compressible turbulent jet.
We have termed \textit{acoustic} resolvent the approach in which the energy of the response in the SVD is optimised in the near-acoustic field, excluding the hydrodynamic region.
Conversely, the \textit{standard} resolvent includes the hydrodynamic domain.
While the SVD of the acoustic resolvent was recently used by different authors \citep{garnaud2013global,jeun2016input,pickering2021resolvent}, the aim of this paper was to shed light on physical interpretation of the SVD of the acoustic resolvent, and assess its potential for jet-noise modelling.
Fundamental differences have been pointed out between the acoustic and standard resolvent analyses.
Whereas the latter exhibits a large gain separation between the first and second singular values, the first singular values of the former slowly decreases before eventually experiencing a quick exponential decay.
Besides, an organised acoustic radiation with a clear directivity is observed in the acoustic resolvent modes.
This directivity is found similar to that of SPOD modes.
This contrasts with the disorganised, scattered acoustic field found in the leading mode of the standard resolvent.
The influence of the nozzle flow has been studied for both configurations.
The gain separation has been found sensitive to the upstream location of the forcing modes in the nozzle.
This is because the first mode is driven by a K-H wavepacket which is unstable in the nozzle and beyond, while sub-optimal modes are associated with the Orr mechanism at play in the jet region.
In the case of the acoustic resolvent, no influence of the nozzle has been reported since acoustic resolvent modes do not promote instability waves.
Instead, the forcing modes contain supersonic waves that radiate Mach waves in the response modes, which is an efficient way to generate acoustic perturbations.
While this complies with the optimality request underpinning the framework of the acoustic resolvent, not all modes appear relevant to jet noise.
The supersonic waves observed in the forcing modes were discussed and  shown to be in line with acoustic matching, from which jet noise is usually understood.
Finally, while the SVD of the acoustic resolvent alone is deemed insufficient to provide a model for jet noise, perspectives were discussed, either by improving this framework \citep{pickering2021resolvent} or by focusing on forcing models \citep{karban2022empirical}.

\section*{Funding}  
This work has received funding from the Clean Sky 2 Joint Undertaking under the European Union's Horizon 2020 research and innovation programme under grant agreement No 78530.
Results reflect only the authors' view and the JU is not responsible for any use that may be made of the information it contains.

\section*{Data Availability Statement}  
Data available on request.

\begin{appendices}

\section{Mesh convergence} \label{ap.mesh}

\begin{figure}
    \centering
        \includegraphics[angle=-90,trim=0 0 0 0, clip,width=0.5\textwidth]{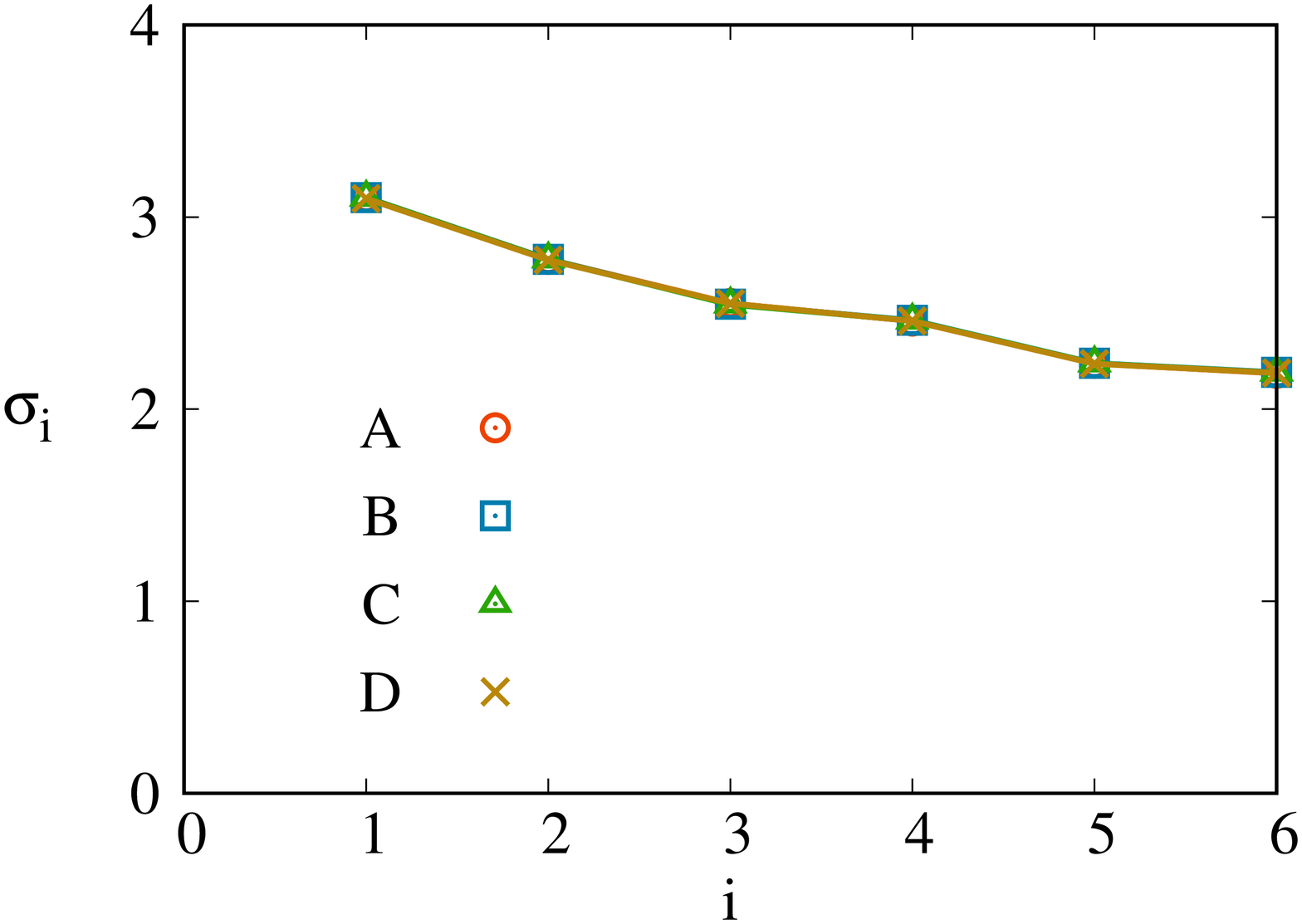}
    \caption{Mesh convergence of the SVD of the acoustic resolvent at $St=0.6$}
     \label{fig.convergence}
\end{figure}

The nomenclature of the different meshes that are used to test the mesh convergence is given in table \ref{tab.meshes}.
Mesh A is the reference mesh used in the paper and is associated with a numerical domain of height $L_r/D=30$.
Finer meshes C and D are tested on a reduced domain $L_r/D=12$.
To check that $L_r$ had no influence on the results, mesh B was introduced by simply truncating mesh A to $L_r/D=12$.
The first six singular values of the acoustic resolvent (reference case) are plotted in figure \ref{fig.convergence}.
This shows identical results for all the meshes tested.

\begin{table}
 \begin{tabular}{lccc}
 \toprule
     Mesh  & $N_x$   &   $N_r$ & $L_r/D$ \\[3pt]
     \midrule
      A   & 595 & 285 & 30\\
      B   & 595 & 229 & 12\\
      C   & 993 & 229 & 12\\
      D   & 595 & 374 & 12\\
      \botrule
 \end{tabular}
 \caption{Nomenclature of the meshes}
 \label{tab.meshes}
\end{table}

\end{appendices}


\bibliography{bibli_turbulent_jet}

\end{document}